\documentclass[aps,prx,superscriptaddress,floatfix,twocolumn,citeautoscript]{revtex4}

\pdfoutput=1 

\usepackage{graphicx}
\usepackage{dcolumn}
\usepackage{bm}
\usepackage{subfigure}
\usepackage{notes2bib}
\usepackage{amsmath} 
\usepackage{amsthm} 
\usepackage{amssymb}	

\let\baraccent=\= 
\renewcommand{\=}[1]{\stackrel{#1}{=}} 

\begin{document}

\title{Exciton fission via ultrafast long-range resonant tunnelling in organic photovoltaic diodes}

\author{Eric R.\ Bittner}
\address{ Department of Chemistry and Physics, University of Houston, Houston, TX 77204}
\address{Department of Physics and
Regroupement qu{\'e}b{\'e}cois sur les mat{\'e}riaux de pointe, Universit{\'e} de Montr{\'e}al,
C.P.\ 6128, Succursale centre-ville, 
Montr{\'e}al (Qu{\'e}bec) H3C~3J7, Canada}

\author{Carlos Silva}
\address{Department of Physics and
Regroupement qu{\'e}b{\'e}cois sur les mat{\'e}riaux de pointe, Universit{\'e} de Montr{\'e}al,
C.P.\ 6128, Succursale centre-ville, 
Montr{\'e}al (Qu{\'e}bec) H3C~3J7, Canada}
\address{Department of Physics (Experimental Solid-State Physics), Imperial College London, South Kensington Campus, London SW7~2AZ, United Kingdom}

\date{\today}

\begin{abstract}
We present an exciton/lattice model of the electronic dynamics of primary photoexcitations in a 
polymeric semiconductor heterojunction 
which includes both polymer $\pi$-stacking, energetic disorder, and phonon relaxation. 
Results from our model are consistent with a wide range of recent 
experimental evidence that excitons decay directly to well-defined polarons 
on a sub-100 fs timescale, which is substantially faster than exciton relaxation processes. 
Averaging over multiple samples,
we find that as the interfacial offset is increased, a substantial fraction of the 
density of electronic states in the energy region about the initial exciton 
carries significant charge-transfer character with two charges separated in the outer regions
of the model lattice.  The results indicate a slight increase in the density of such current-producing 
states if the region close to the interface is more disordered.   However, 
since their density of states overlaps the excitation line-shape of the 
primary exciton, we show that it is possible that the exciton can decay directly into current-producing 
states via tunneling on an ultrafast time-scale.  We find  this process to be
 {\em independent} of the location of energetic disorder in the 
system, and hence we expect exciton fission via resonant tunnelling  to be a ubiquitous feature of these systems.

\vspace{10pt}
Subject Areas: Chemical Physics, Condensed Matter Physics, Materials Science

\end{abstract}


\maketitle
\section{Introduction}

Photovoltaic diodes based on blends of semiconductor polymers and fullerene derivatives now produce power conversion efficiencies exceeding 10\% under standard solar illumination~\cite{He:2012fk}. This indicates that photocarriers can be generated efficiently in well-optimized organic heterostructures, but the mechanism for converting highly-bound Frenkel excitons to photocarriers is not understood in spite of vigorous, multidisciplinary research activity. 
A detailed mechanistic understanding of primary charge generation dynamics is of key fundamental importance in the development of organic solar cells, and we propose it to be generally important in photoinduced charge-transfer processes in condensed matter.  
Recent spectroscopic measurements on organic photovoltaic systems have reported that charged photoexcitations 
can be generated on $\leq 100$-fs timescales~\cite{Sariciftci:1994kx,Banerji:2010vn,Tong2010,Sheng2012,Jailaubekov:2013fk,Grancini:2013uq,doi:10.1021/jz4010569,doi:10.1021/jp4071086,Banerji:2013ej}
 but full charge separation to produce photocarriers is expected to be energetically expensive given strong Coulombic barriers due to the low dielectric constant in molecular semiconductors. Nonetheless,  experiments  by G\'elinas {\em et~al.}, in which Stark-effect signatures in transient absorption spectra were analyzed to probe the local electric field as charge separation proceeds, indicate that electrons and holes separate by $\sim40$\AA\ over the first 100\,fs\ and evolve further on picosecond timescales to produce unbound charge pairs~\cite{Gelinas:2013fk}. Concurrently, Provencher {\em et~al.} have demonstrated, via transient resonance-Raman measurements, clear polaronic vibrational signatures on sub-100-fs on the polymer backbone, with very limited molecular reorganization or vibrational relaxation following the ultrafast step~\cite{Provencher:fk}. Such spectacularly and apparently universally rapid through-space charge transfer between excitons on the polymer backbone and acceptors across the heterojunction would be difficult to rationalize within Marcus theory within a localized basis without invoking unphysical distance dependence of tunnelling rate constants~\cite{Barbara:1996uc}.
We  focus here on the role that quantum coherence dynamics in a disordered $\pi$-stacked polymer lattice, which are correlated to the 
dynamical motion of the molecular framework~\cite{Rozzi:2013fk}, and that of energetic  disorder in 
promoting the asymptotic separation of mobile charge pairs following photoexcitation. The significant element in the context of ultrafast charge separation in the system considered 
here is the involvement of delocalized charge-transfer states in the early quantum dynamics of the exciton. 
Our model is based upon a Frenkel exciton lattice model parameterized to describe the 
$\pi$-electronic states of a polymeric-semiconductor type-II heterojunction~\cite{karabunarliev:3988,karabunarliev:4291,karabunarliev:10219,karabunarliev:057402}.  

We introduce energetic disorder by randomizing the local site energies 
in the region about the heterojunction interface or in the region away from the interface and 
analyze the resulting electron-hole eigenstates.   
Polymer microstructural probes have 
revealed general relationships between disorder, aggregation and electronic properties in polymeric semiconductors~\cite{Noriega:2013uq}.   Moreover, aggregation (ordering) can be perturbed by 
varying the blend-ratio and composition of donor and acceptor polymers~\cite{doi:10.1021/jp104111h}.
 On one hand, energetic disorder at the interface would provide a
free-energy gradient for localized charge-transfer states to escape to the 
asymptotic regions.  In essence, the localized polarons in the interfacial region
could escape into band of highly mobile polarons away from the heterojunction region~\cite{doi:10.1021/nl302172w}.  On the other, 
energetic disorder in the regions away from the interface would provide an entropic 
driving force by increasing the density of localized polaron states
away from the interfacial region, allowing the polarons to hop or diffuse away from the 
interface before recombination could take place~\cite{doi:10.1021/jp2083133}.
Finally, a report by Bakulin~{\em et~al.} indicates that if relaxed charge-transfer-excitons are pushed with an infrared pulse,
 they increase photocurrent via delocalized states rather than by energy gradient-driven 
hopping~\cite{Bakulin16032012}.


\begin{figure*}[t]
\subfigure[]{\includegraphics[width=8cm]{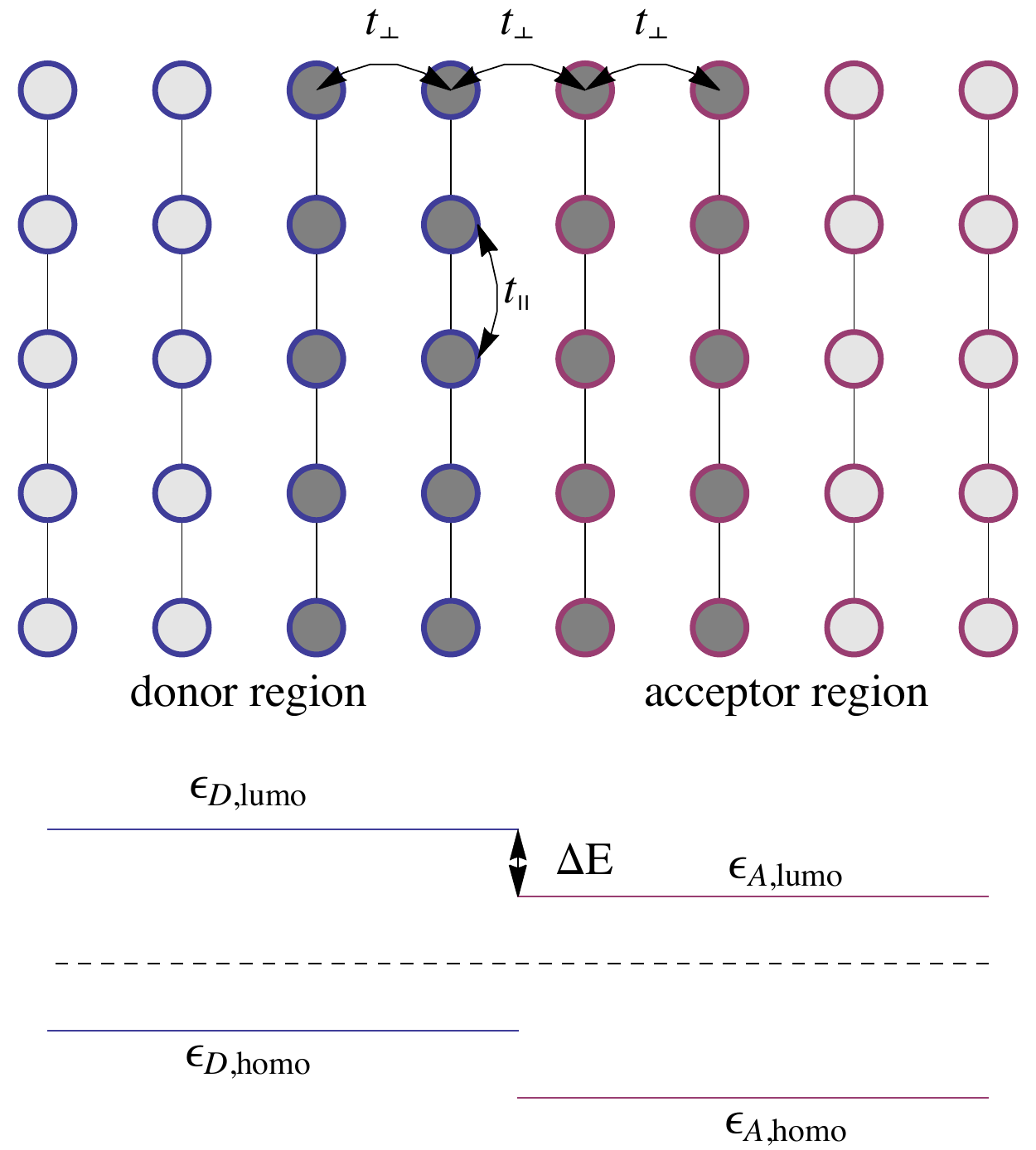}}
\subfigure[]{\includegraphics[width=8cm]{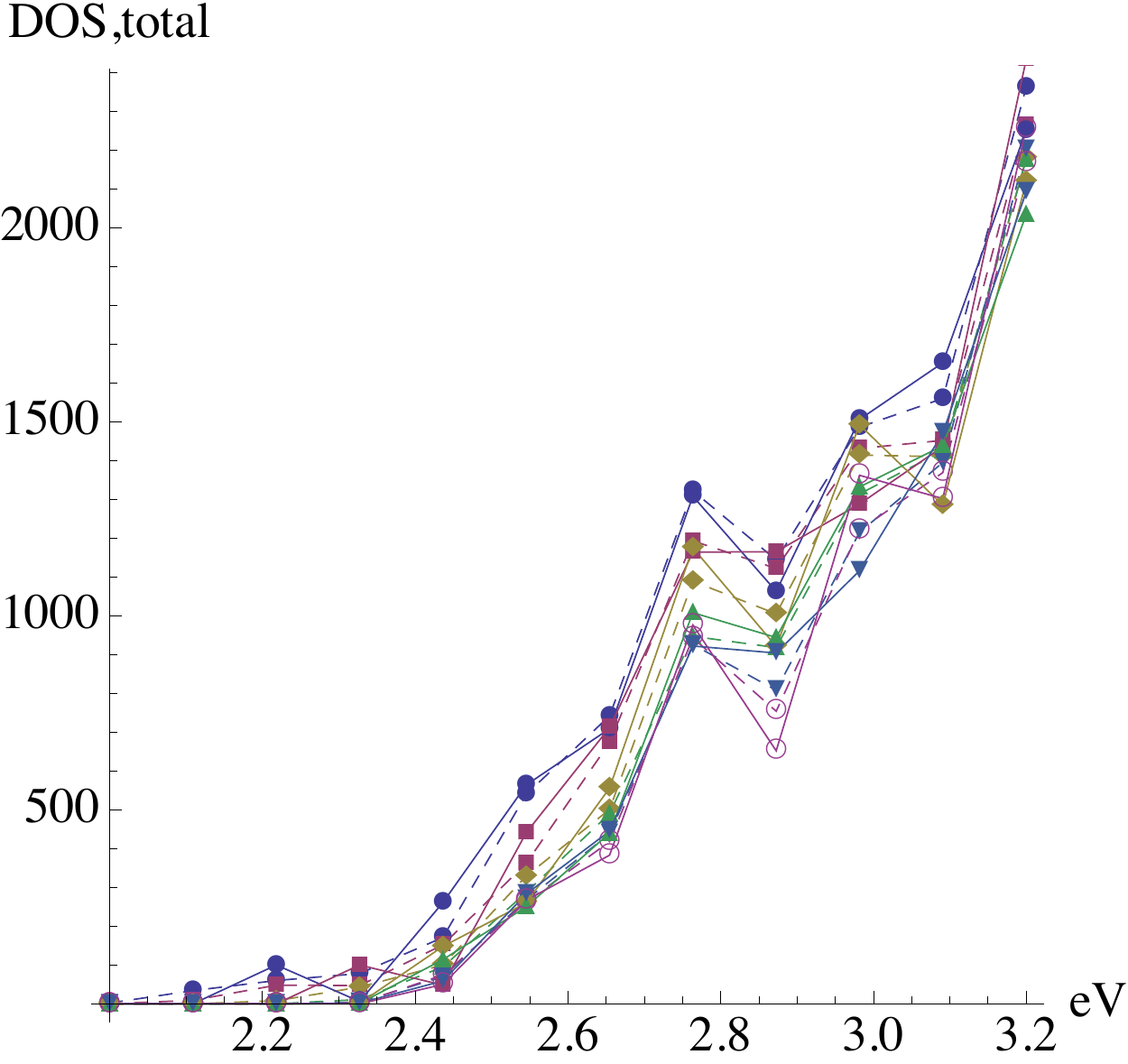}}
\caption{(a) Sketch of lattice model showing parallel aligned chains and location of heterojunction boundary.  Lower inset
shows the relative energy offset between donor and acceptor bands.  Intrachain ($t_\parallel$) and interchain ($t_\perp$) hopping
integrals are also indicated. (b)  Density of electronic excited states for both ODO (dashed) and DOD  (solid) for 
increasing interfacial bias. (purple: $\Delta E = 0$ to blue: $\Delta E = 0.5eV$.)}\label{dos}
\end{figure*}

\section{Lattice exciton model}
We employ a two-band exciton/phonon model we described previously modified as 
to include both stacking and a band-offset to define a heterojunction domain~\cite{karabunarliev:3988,karabunarliev:4291,karabunarliev:10219,karabunarliev:057402}.  
Each site contributes a valance and conduction band Wannier orbital and the ground state consists
of each site being doubly occupied.    
A sketch of our model system is shown  Fig.~\ref{dos}a. 
The blue sites have an energy band off-set relative to the red-sites of $\Delta E $ which provides the necessary driving force for charge-separation at the interface between the two domains. 
The important distinction between the present model and 
that for a single chain is that we estimate the interchain hopping term is an order of magnitude 
smaller than the intrachain hopping term $t_{\perp} \approx t_{\parallel}/10$.  
Single electron/hole excitations from the ground-state are considered within 
configuration interaction (CI) theory which includes all singlet electron/hole configurations.

 \begin{figure*}[t]
\subfigure[]{\includegraphics[width=8.0cm]{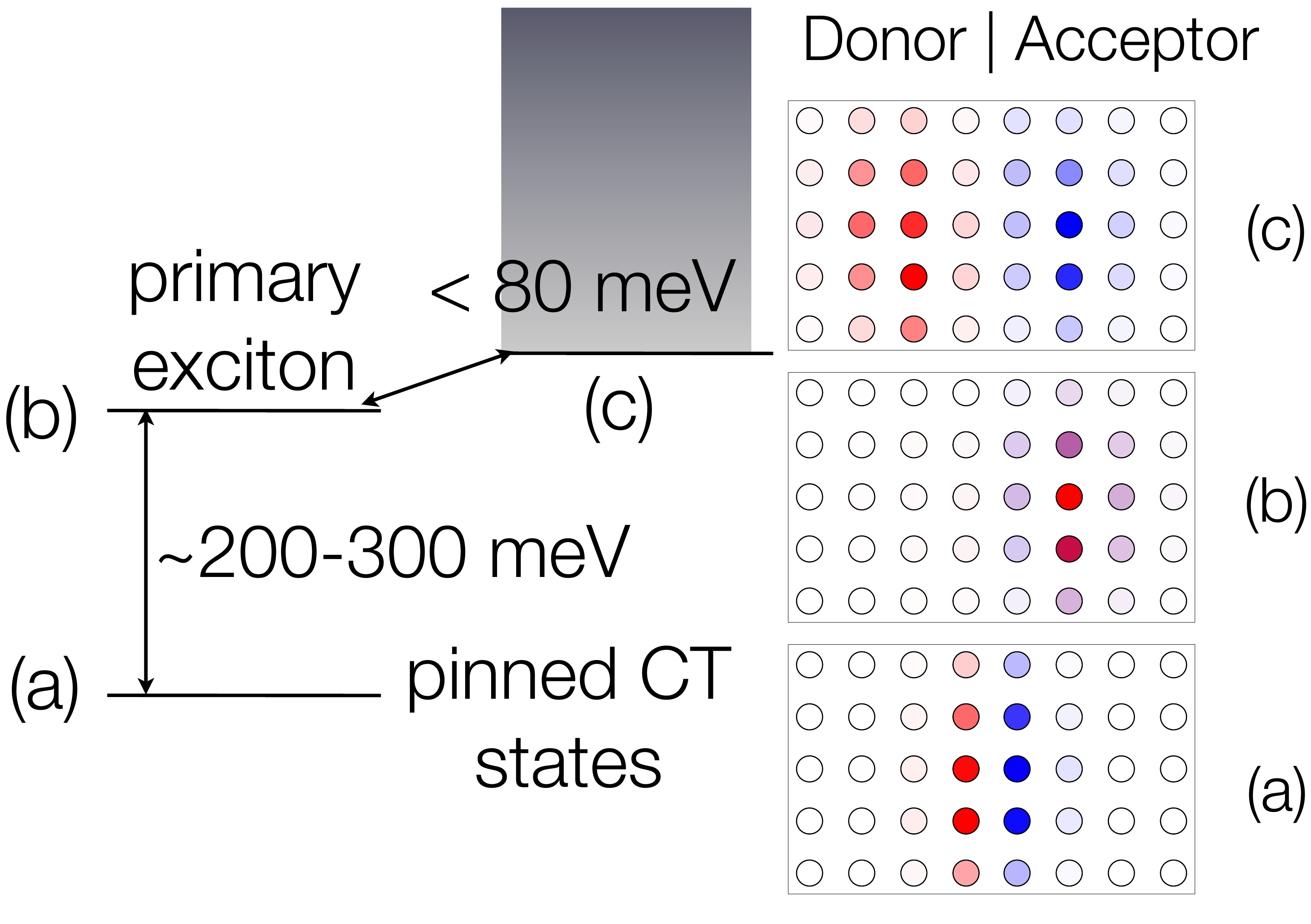}}
\subfigure[]{\includegraphics[width=8.0cm]{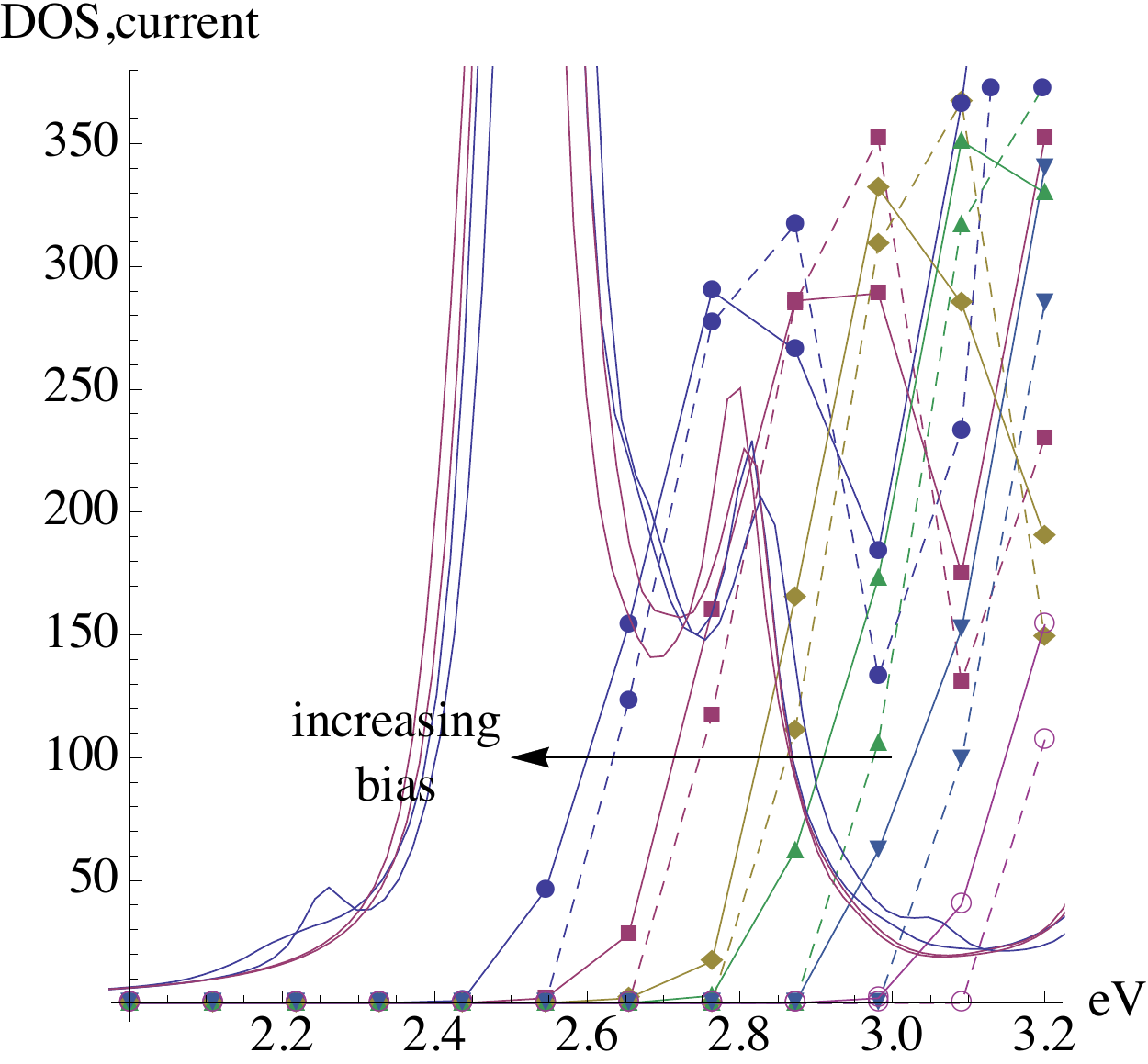}}
\caption{
(a) Relative energy levels and electron/hole densities for selected low-lying states.
(b) Computed absorption profiles (solid blue and red)  and density of current producing states (lines/points)
for increasing interfacial bias. 
Blue absorption peaks correspond to the $\Delta E =0.5$ eV off-set cases and the red correspond to the $\Delta E= 0$ eV
lattice. Both ODO and DOD cases are shown.  The curves with data points correspond to the 
density of current producing states with increasing bias from $\Delta E = 0$ to 0.5 eV. 
 Dashed = ODO and solid = DOD. }\label{Scheme}
\end{figure*}

%
Each local site also contributes  two localized  phonon modes which 
modulate the energy gap at each site and give rise to 
optical phonon bands that are delocalized across each chain. No explicit interchain 
vibronic coupling was included in our model.  The two bands included correspond to the 
Franck-Condon active modes that contribute to the absorption and fluorescence spectra of 
conjugated polymer containing phenylene groups such as poly-pheneylene-vinylene (PPV).
Specifically, these correspond to low frequency torsional modes with frequencies ???
 and the higher frequency C=C stretching modes which are well-known to be 
 coupled to the $\pi$-electronic states in poly-phenylene type conjugated polymer systems.
 With exception as noted above, the electronic couplings and electron/phonon couplings used here 
 are the same as reported in our earlier works on 
 exciton dissociation at interfaces~\cite{Tamura:2006,tamura:021103,tamura:107402}.  A complete listing of
 the interaction terms and parameters is given in Ref.~\citenum{karabunarliev:4291}.

\subsection{The role of interfacial disorder}

To explore the role of interfacial disorder, we include a small amount of 
gaussian noise ($\sigma/t_\parallel = 0.2$)
in the local HOMO--LUMO gaps in either the region close to the interface or 
in the regions away from the interface.  
We shall refer to the two as  the ``ordered-disordered-ordered'' model  (ODO)
and ``disordered-ordered-disordered'' model (DOD) respectively,
In Fig.~\ref{dos}b  we show the average density of states produced by our model as the energy off-set 
between the donor and acceptor region is increased. By and large, the total density of 
states produced by the ODO  and DOD models  are indistinguishable.  Increasing the interfacial bias
generally produces a shift towards lower energy states composed primarily of charge-transfer 
states pinned to the donor-acceptor interface.  The averaged density of states appears to be 
insensitive to whether or not the interface is disordered or ordered within the context of our model.

In the presence of an energy off-set at the interface, the lowest energy single-excitations
are composed of electron/hole configurations pinned along the interface (i.e. exciplex states).
Even though charge separation has occurred, little or no photocurrent 
can be produced since the binding energy of the exciplex is greater than the thermal energy.
In order to produce current, the electron and hole must overcome their mutual 
Coulombic attraction and be separated at a distance $R > 2e^{2}/(3\epsilon k_{B}T)$
(assuming a homogeneous 3D medium with dielectric constant $\epsilon$).   Within our 
model we define current-producing states as any elementary configuration with 
the positive hole on the outer-most donor chain and the negative electron 
on the outer-most acceptor chain.   Such states serve as gateway states for charges to escape 
from the local heterojunction region.\cite{bittner:034707}

Fig.~\ref{Scheme} shows three types of singlet  states produced by our ODO lattice model with $\Delta E = 0.5$eV 
 with their relative energetic ordering.   Shown are the charge densities on each site.  
 State (a) typifies the lowest energy excited state of our model with positive charge (hole density) shown in red 
 and negative charge (electron density) shown in blue.
The lowest energy state produced by our model is an interracially pinned charge-transfer
 state that is delocalized over the entire length of the interface.  Such states are typically 300meV lower in energy
 than the primary exciton shown in (b) and carry some oscillator strength to the ground-state due to 
 mixing between purely charge-transfer and neutral electron/hole configurations and may be 
 referred to as a charge-transfer exciton or an exciplex.
This excitonic state (b) is primarily composed of neutral electron/hole configurations
 and is delocalized over three of the polymer chains on the ``acceptor" side of the heterojunction.  
State (c)  typifies the charge-transfer/polaron states that lie very close in energy to the primary exciton.  
In the case shown here, the hole (red) has been transferred across the heterojunction and is separated from the 
electronic by 3-5 polymer chains.  Such states lie 80-100 meV above the primary exciton.
The oscillations in the electron/hole density indicate that the state has kinetic energy perpendicular to the interface
 and as such are likely channels for direct fission of the excition into states capable of producing photocurrent.

in Fig.~\ref{Scheme}b  we show the absorption line-shapes for both the ODO and DOD cases
along with the density of ``current producing'' states as defined by requiring the 
states to have a threshold density of 10\% of their electron/hole configurations on two outermost 
polymer chains corresponding to states with maximal charge separation.  Clearly, 
increasing the interfacial bias shifts the fraction of states capable of directly producing 
current towards lower energies to the extent that they begin to overlap the absorption 
spectrum of the system.  It should also be noted that the energetic onset
 for producing current is somewhat higher for the DOD cases suggesting that
disorder at the interfacial region may facilitate
 efficient charge separation as suggested by Forrest.
\cite{peumans:158,doi:10.1021/nl302172w}

\subsection{Couplings due to phonon fluctuations}   
We now consider the  couplings and state-to-state transition rates between the 
primary exciton and its neighboring states due to fluctuations and noise in the 
phonon degrees of freedom included our model.   For this, we first obtain the 
full set of diagonal  and off-diagonal exciton/phonon couplings and then transform 
the Hamiltonian into a dressed representation using a non-adiabatic polaron transformation
described by Pereverzev and Bittner.\cite{pereverzev:104906}   We refer the reader to Ref.~\cite{pereverzev:104906} 
for explicit derivations and expressions.
We then use this to construct time correlation functions of the exciton/phonon coupling operator
and compute the spectral density {\em viz.}
$$
S_{mn}(\omega)=\int_{-\infty}^\infty dt\langle \hat V_{nm}(-t)\hat V_{mn}(t)\rangle
e^{-i\omega t}, \label{rates}
$$
where $\hat V_{nm}(t)$ is the polaron-transformed electron/phonon operator written in the Heisenberg representation 
and $\langle \cdots \rangle$ is the thermal average over phonon degrees of freedom
\begin{eqnarray}
\langle V_{nm}V_{mn}(t)\rangle={\rm Tr}\left[
\sum_{ij}
\left(V_{nmi}V_{mnj}(t)\rho^{os}_{eq}\right)\right].\label{corrfun}
\end{eqnarray} 
and $\rho^{o/c}_{eq}$ is the equilibrium density operator for the oscillator degrees of freedom.
It is important to note that in the perturbative regime, $S_{nm}(\omega_{nm})$ is the Fermi's ``golden rule'' transition rate between states $n\to m$.
Integrating the spectral density over frequencies gives a phonon-averaged coupling between states
\begin{eqnarray}
W^2 = \int \frac{d\omega}{2 \pi}S(\omega).
\end{eqnarray}
When $\sqrt{W^2}$ is comparable to the energy gap between states $n$ and $m$, phonon fluctuations can induce 
resonant tunneling between the two states.  
\cite{Bittner:2013aa}
 
 \begin{figure*}[t]
\subfigure[]{\includegraphics[width=8.0cm]{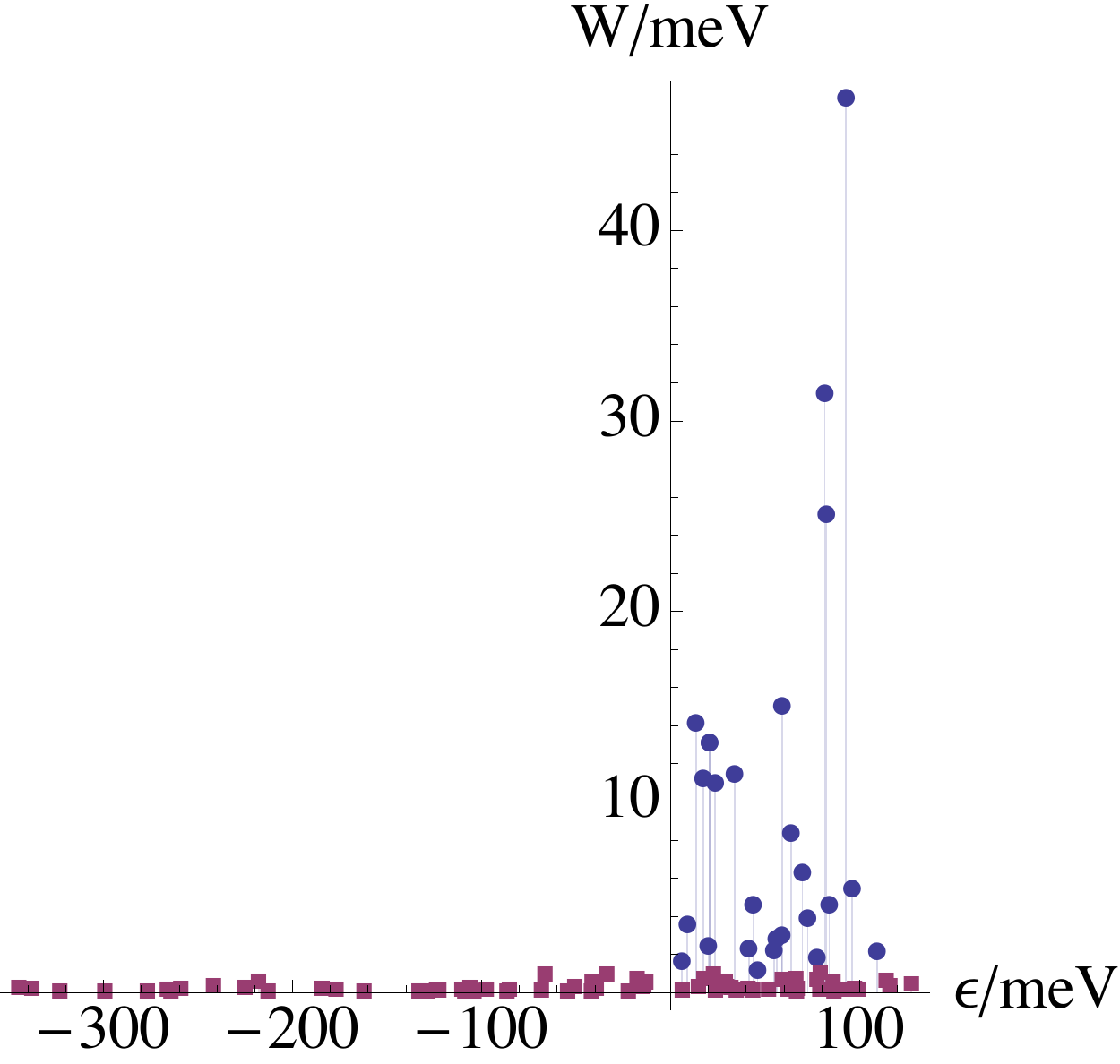}}
\subfigure[]{\includegraphics[width=8.0cm]{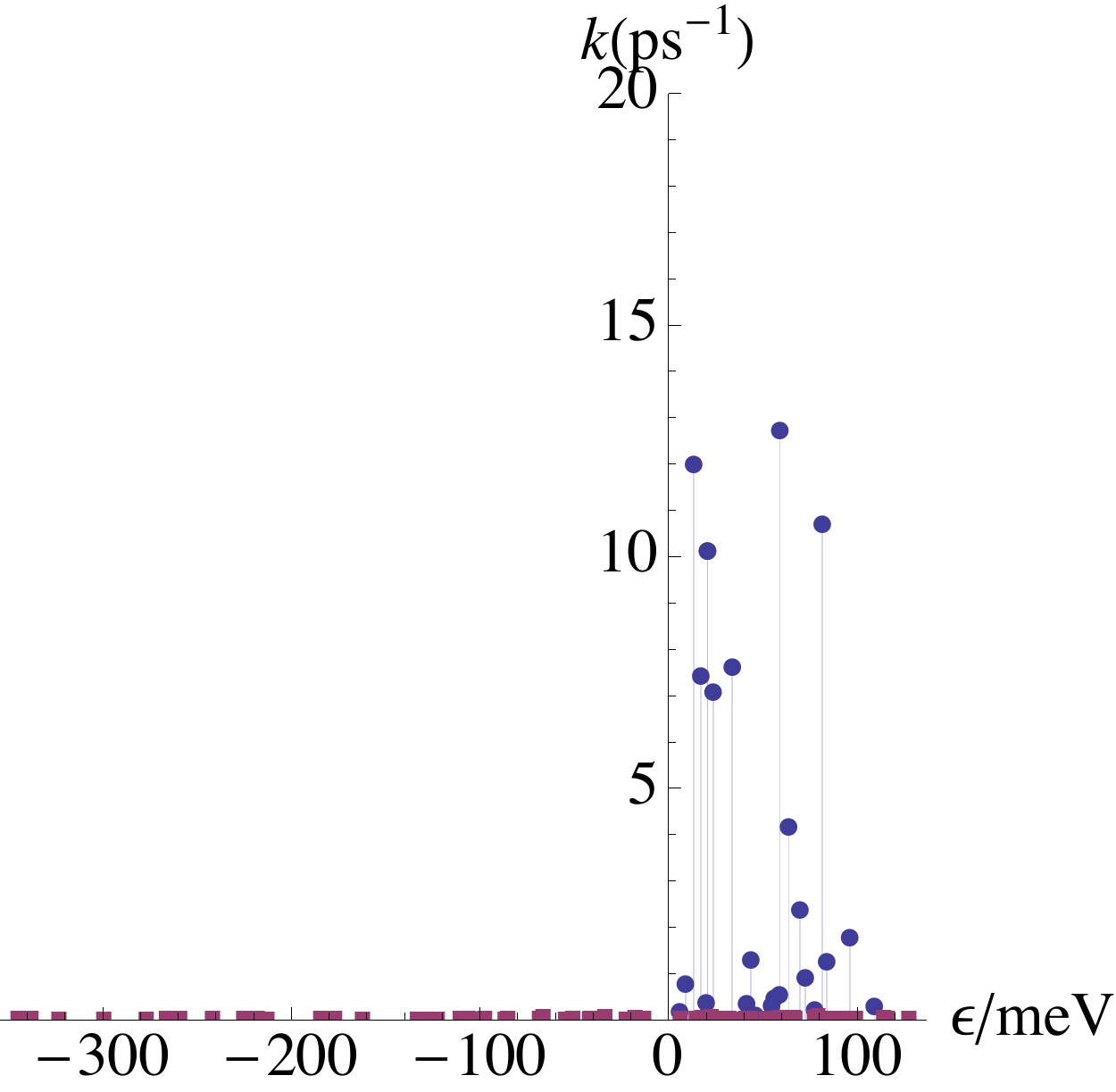}}
\caption{(a) Distribution of couplings and (b) golden-rule rates between primary exciton and nearby states at 10K.
Blue points correspond to transition from the primary exciton to current producing states.
}\label{coupling}
\end{figure*}

In Figure~\ref{coupling}a we show the distribution of noise-averaged interstate
 couplings between the primary exciton and each of its neighboring states for the ODO polymer model with a 
$\Delta E = 0.5$eV offset (sampled over multiple realizations of the lattice).  States with $E<E_{ex}$ are predominantly interfacially
bound electron-hole pairs (exciplex states).   
On average, these lower-lying states are coupled very weakly  to the primary exciton
indicating that relaxation to the lower-lying pinned CT state proceeds via a sequence of 
micro-states.  
It is surprising, however, that in many cases, the  coupling between the exciton to states
 that are immediately higher in energy are
 on the order of the energy difference between the primary exciton and the final state.
   This puts initial (primary exciton) and 
 final state into the strong coupling or resonant tunneling regime.   Since the majority
 of the states with large couplings correspond to current-producing states as defined above, 
 Comparing this plot to the density of current producing states 
one can conclude that the primary exciton can decay {\em directly} into the current producing 
 states by resonant tunneling. 
 
 In Fig.\ref{coupling}b we show the distribution of golden-rule rate constants for transitions originating from 
 the primary exciton (again sampled over multiple realizations of the lattice). 
 defining $\overline{k}_+$ as the average rate to all nearby higher energy states and $\overline{k}_-$ as the average rate to 
 all lower energy states, we can estimate the time-scale for exciton fission into current producing states
 as $1/\overline{k}_+ = 35$\,fs and the time-scale for exciton fission into pinned charge transfer states as $1/\overline{k}_- = 43$\,ps.
 Both of these estimates are consistent with a wide-range of recent experimental evidence indicating that 
 excitons decay into current producing states on time-scales of $<100 $\,fs.  
 
\section{Conclusion}
We presented here a lattice model for the interfacial electronic states of a polymer-polymer bulk-heterojunction interface.
With increasing band off-set, a sizable density of ``gate-way" states is brought into a
resonance with the primary photo-excitation.  As the energy difference decreases couplings due to phonon fluctuations 
become equivalent the energy difference between the states, leading to 
long-range resonant tunneling of charges away from the initial excitation
 on an average time-scale of 35 fs.  Taking the spacing between polymer chains to be 
 $\approx 4$\,\AA\,  over this timescale, the electron or hole would tunnel 12-15\,\AA\ away from the 
site of the initial excitation within 35\,fs following excitation. 
Moreover, given the universal nature of our model, 
 we propose that that direct resonant tunneling to current producing states plays a ubiquitous role in the ultrafast
 dissociation of excitons in organic  BHJ cells.

\begin{acknowledgments}
We thank Simon G\'elinas and Sir Richard Friend for sharing Ref.~\cite{Gelinas:2013fk}
 prior to publication. 
We also thank Prof. Irene Burghardt for discussions at the on-set of this work.
The work at the University of Houston was funded in part by the National Science Foundation 
(CHE-1011894) and the Robert A. Welch Foundation (E-1334).
CS acknowledges support from the Canada Research Chair in Organic Semiconductor Materials.
ERB acknowledges support from  Fulbright Canada and the US Department of State.  
\end{acknowledgments}


\begin{thebibliography}{30}
\expandafter\ifx\csname natexlab\endcsname\relax\def\natexlab#1{#1}\fi
\expandafter\ifx\csname bibnamefont\endcsname\relax
  \def\bibnamefont#1{#1}\fi
\expandafter\ifx\csname bibfnamefont\endcsname\relax
  \def\bibfnamefont#1{#1}\fi
\expandafter\ifx\csname citenamefont\endcsname\relax
  \def\citenamefont#1{#1}\fi
\expandafter\ifx\csname url\endcsname\relax
  \def\url#1{\texttt{#1}}\fi
\expandafter\ifx\csname urlprefix\endcsname\relax\def\urlprefix{URL }\fi
\providecommand{\bibinfo}[2]{#2}
\providecommand{\eprint}[2][]{\url{#2}}

\bibitem[{\citenamefont{He et~al.}(2012)\citenamefont{He, Zhong, Su, Xu, Wu,
  and Cao}}]{He:2012fk}
\bibinfo{author}{\bibfnamefont{Z.}~\bibnamefont{He}},
  \bibinfo{author}{\bibfnamefont{C.}~\bibnamefont{Zhong}},
  \bibinfo{author}{\bibfnamefont{S.}~\bibnamefont{Su}},
  \bibinfo{author}{\bibfnamefont{M.}~\bibnamefont{Xu}},
  \bibinfo{author}{\bibfnamefont{H.}~\bibnamefont{Wu}}, \bibnamefont{and}
  \bibinfo{author}{\bibfnamefont{Y.}~\bibnamefont{Cao}}, \bibinfo{journal}{Nat
  Photon} \textbf{\bibinfo{volume}{6}}, \bibinfo{pages}{591}
  (\bibinfo{year}{2012}),
  \urlprefix\url{http://dx.doi.org/10.1038/nphoton.2012.190}.

\bibitem[{\citenamefont{Sariciftci and Heeger}(1994)}]{Sariciftci:1994kx}
\bibinfo{author}{\bibfnamefont{N.}~\bibnamefont{Sariciftci}} \bibnamefont{and}
  \bibinfo{author}{\bibfnamefont{A.}~\bibnamefont{Heeger}},
  \bibinfo{journal}{Int. J. Mod. Phys. B} \textbf{\bibinfo{volume}{8}},
  \bibinfo{pages}{237} (\bibinfo{year}{1994}).

\bibitem[{\citenamefont{Banerji et~al.}(2010)\citenamefont{Banerji, Cowan,
  Leclerc, Vauthey, and Heeger}}]{Banerji:2010vn}
\bibinfo{author}{\bibfnamefont{N.}~\bibnamefont{Banerji}},
  \bibinfo{author}{\bibfnamefont{S.}~\bibnamefont{Cowan}},
  \bibinfo{author}{\bibfnamefont{M.}~\bibnamefont{Leclerc}},
  \bibinfo{author}{\bibfnamefont{E.}~\bibnamefont{Vauthey}}, \bibnamefont{and}
  \bibinfo{author}{\bibfnamefont{A.~J.} \bibnamefont{Heeger}},
  \bibinfo{journal}{J. Am. Chem. Soc.} \textbf{\bibinfo{volume}{132}},
  \bibinfo{pages}{17459} (\bibinfo{year}{2010}).

\bibitem[{\citenamefont{Tong et~al.}(2010)\citenamefont{Tong, Coates, Moses,
  Heeger, Beaupr{\'e}, and Leclerc}}]{Tong2010}
\bibinfo{author}{\bibfnamefont{M.}~\bibnamefont{Tong}},
  \bibinfo{author}{\bibfnamefont{N.}~\bibnamefont{Coates}},
  \bibinfo{author}{\bibfnamefont{D.}~\bibnamefont{Moses}},
  \bibinfo{author}{\bibfnamefont{A.~J.} \bibnamefont{Heeger}},
  \bibinfo{author}{\bibfnamefont{S.}~\bibnamefont{Beaupr{\'e}}},
  \bibnamefont{and} \bibinfo{author}{\bibfnamefont{M.}~\bibnamefont{Leclerc}},
  \bibinfo{journal}{Phys. Rev. B} \textbf{\bibinfo{volume}{81}},
  \bibinfo{pages}{125210} (\bibinfo{year}{2010}).

\bibitem[{\citenamefont{Sheng et~al.}(2012)\citenamefont{Sheng, Basel, Pandit,
  and Vardeny}}]{Sheng2012}
\bibinfo{author}{\bibfnamefont{C.-X.} \bibnamefont{Sheng}},
  \bibinfo{author}{\bibfnamefont{T.}~\bibnamefont{Basel}},
  \bibinfo{author}{\bibfnamefont{B.}~\bibnamefont{Pandit}}, \bibnamefont{and}
  \bibinfo{author}{\bibfnamefont{Z.~V.} \bibnamefont{Vardeny}},
  \bibinfo{journal}{Organic Electronics} \textbf{\bibinfo{volume}{13}},
  \bibinfo{pages}{1031} (\bibinfo{year}{2012}).

\bibitem[{\citenamefont{Jailaubekov et~al.}(2013)\citenamefont{Jailaubekov,
  Willard, Tritsch, Chan, Sai, Gearba, Kaake, Williams, Leung, Rossky
  et~al.}}]{Jailaubekov:2013fk}
\bibinfo{author}{\bibfnamefont{A.~E.} \bibnamefont{Jailaubekov}},
  \bibinfo{author}{\bibfnamefont{A.~P.} \bibnamefont{Willard}},
  \bibinfo{author}{\bibfnamefont{J.~R.} \bibnamefont{Tritsch}},
  \bibinfo{author}{\bibfnamefont{W.-L.} \bibnamefont{Chan}},
  \bibinfo{author}{\bibfnamefont{N.}~\bibnamefont{Sai}},
  \bibinfo{author}{\bibfnamefont{R.}~\bibnamefont{Gearba}},
  \bibinfo{author}{\bibfnamefont{L.~G.} \bibnamefont{Kaake}},
  \bibinfo{author}{\bibfnamefont{K.~J.} \bibnamefont{Williams}},
  \bibinfo{author}{\bibfnamefont{K.}~\bibnamefont{Leung}},
  \bibinfo{author}{\bibfnamefont{P.~J.} \bibnamefont{Rossky}},
  \bibnamefont{et~al.}, \bibinfo{journal}{Nat Mater}
  \textbf{\bibinfo{volume}{12}}, \bibinfo{pages}{66} (\bibinfo{year}{2013}),
  \urlprefix\url{http://dx.doi.org/10.1038/nmat3500}.

\bibitem[{\citenamefont{Grancini et~al.}(2013)\citenamefont{Grancini, Maiuri,
  Fazzi, Petrozza, Egelhaaf, Brida, Cerullo, and Lanzani}}]{Grancini:2013uq}
\bibinfo{author}{\bibfnamefont{G.}~\bibnamefont{Grancini}},
  \bibinfo{author}{\bibfnamefont{M.}~\bibnamefont{Maiuri}},
  \bibinfo{author}{\bibfnamefont{D.}~\bibnamefont{Fazzi}},
  \bibinfo{author}{\bibfnamefont{A.}~\bibnamefont{Petrozza}},
  \bibinfo{author}{\bibfnamefont{H.-J.} \bibnamefont{Egelhaaf}},
  \bibinfo{author}{\bibfnamefont{D.}~\bibnamefont{Brida}},
  \bibinfo{author}{\bibfnamefont{G.}~\bibnamefont{Cerullo}}, \bibnamefont{and}
  \bibinfo{author}{\bibfnamefont{G.}~\bibnamefont{Lanzani}},
  \bibinfo{journal}{Nat Mater} \textbf{\bibinfo{volume}{12}},
  \bibinfo{pages}{29} (\bibinfo{year}{2013}),
  \urlprefix\url{http://dx.doi.org/10.1038/nmat3502}.

\bibitem[{\citenamefont{Kaake et~al.}(2013)\citenamefont{Kaake, Moses, and
  Heeger}}]{doi:10.1021/jz4010569}
\bibinfo{author}{\bibfnamefont{L.~G.} \bibnamefont{Kaake}},
  \bibinfo{author}{\bibfnamefont{D.}~\bibnamefont{Moses}}, \bibnamefont{and}
  \bibinfo{author}{\bibfnamefont{A.~J.} \bibnamefont{Heeger}},
  \bibinfo{journal}{The Journal of Physical Chemistry Letters}
  \textbf{\bibinfo{volume}{4}}, \bibinfo{pages}{2264} (\bibinfo{year}{2013}),
  \eprint{http://pubs.acs.org/doi/pdf/10.1021/jz4010569},
  \urlprefix\url{http://pubs.acs.org/doi/abs/10.1021/jz4010569}.

\bibitem[{\citenamefont{Mukamel}(2013)}]{doi:10.1021/jp4071086}
\bibinfo{author}{\bibfnamefont{S.}~\bibnamefont{Mukamel}},
  \bibinfo{journal}{The Journal of Physical Chemistry A}
  \textbf{\bibinfo{volume}{in press}} (\bibinfo{year}{2013}),
  \eprint{http://pubs.acs.org/doi/pdf/10.1021/jp4071086},
  \urlprefix\url{http://pubs.acs.org/doi/abs/10.1021/jp4071086}.

\bibitem[{\citenamefont{Banerji}(2013)}]{Banerji:2013ej}
\bibinfo{author}{\bibfnamefont{N.}~\bibnamefont{Banerji}}, \bibinfo{journal}{J.
  Mater. Chem. C} \textbf{\bibinfo{volume}{1}}, \bibinfo{pages}{3052}
  (\bibinfo{year}{2013}).

\bibitem[{\citenamefont{G{\'e}linas
  et~al.}(Submitted)\citenamefont{G{\'e}linas, Rao, Kumar, Smith, Chin, Clark,
  van~der Poll, Bazan, and Friend}}]{Gelinas:2013fk}
\bibinfo{author}{\bibfnamefont{S.}~\bibnamefont{G{\'e}linas}},
  \bibinfo{author}{\bibfnamefont{A.}~\bibnamefont{Rao}},
  \bibinfo{author}{\bibfnamefont{A.}~\bibnamefont{Kumar}},
  \bibinfo{author}{\bibfnamefont{S.~L.} \bibnamefont{Smith}},
  \bibinfo{author}{\bibfnamefont{A.~W.} \bibnamefont{Chin}},
  \bibinfo{author}{\bibfnamefont{J.}~\bibnamefont{Clark}},
  \bibinfo{author}{\bibfnamefont{T.~S.} \bibnamefont{van~der Poll}},
  \bibinfo{author}{\bibfnamefont{G.~C.} \bibnamefont{Bazan}}, \bibnamefont{and}
  \bibinfo{author}{\bibfnamefont{R.~H.} \bibnamefont{Friend}},
  \bibinfo{journal}{Science}  (\bibinfo{year}{Submitted}).

\bibitem[{\citenamefont{Provencher et~al.}(2013)\citenamefont{Provencher,
  B\'erub\'e, Parker, Greetham, Towrie, Hellmann, C\^{o}t\'e, Stingelin, Silva,
  and Hayes}}]{Provencher:fk}
\bibinfo{author}{\bibfnamefont{F.}~\bibnamefont{Provencher}},
  \bibinfo{author}{\bibfnamefont{N.}~\bibnamefont{B\'erub\'e}},
  \bibinfo{author}{\bibfnamefont{A.~W.} \bibnamefont{Parker}},
  \bibinfo{author}{\bibfnamefont{G.~M.} \bibnamefont{Greetham}},
  \bibinfo{author}{\bibfnamefont{M.}~\bibnamefont{Towrie}},
  \bibinfo{author}{\bibfnamefont{C.}~\bibnamefont{Hellmann}},
  \bibinfo{author}{\bibfnamefont{M.}~\bibnamefont{C\^{o}t\'e}},
  \bibinfo{author}{\bibfnamefont{N.}~\bibnamefont{Stingelin}},
  \bibinfo{author}{\bibfnamefont{C.}~\bibnamefont{Silva}}, \bibnamefont{and}
  \bibinfo{author}{\bibfnamefont{S.~C.} \bibnamefont{Hayes}},
  \bibinfo{journal}{Nat Mater} \textbf{\bibinfo{volume}{Submitted}}
  (\bibinfo{year}{2013}).

\bibitem[{\citenamefont{Barbara et~al.}(1996)\citenamefont{Barbara, Meyer, and
  Ratner}}]{Barbara:1996uc}
\bibinfo{author}{\bibfnamefont{P.}~\bibnamefont{Barbara}},
  \bibinfo{author}{\bibfnamefont{T.}~\bibnamefont{Meyer}}, \bibnamefont{and}
  \bibinfo{author}{\bibfnamefont{M.}~\bibnamefont{Ratner}}, \bibinfo{journal}{J
  Phys Chem-Us} \textbf{\bibinfo{volume}{100}}, \bibinfo{pages}{13148}
  (\bibinfo{year}{1996}).

\bibitem[{\citenamefont{Andrea~Rozzi et~al.}(2013)\citenamefont{Andrea~Rozzi,
  Maria~Falke, Spallanzani, Rubio, Molinari, Brida, Maiuri, Cerullo, Schramm,
  Christoffers et~al.}}]{Rozzi:2013fk}
\bibinfo{author}{\bibfnamefont{C.}~\bibnamefont{Andrea~Rozzi}},
  \bibinfo{author}{\bibfnamefont{S.}~\bibnamefont{Maria~Falke}},
  \bibinfo{author}{\bibfnamefont{N.}~\bibnamefont{Spallanzani}},
  \bibinfo{author}{\bibfnamefont{A.}~\bibnamefont{Rubio}},
  \bibinfo{author}{\bibfnamefont{E.}~\bibnamefont{Molinari}},
  \bibinfo{author}{\bibfnamefont{D.}~\bibnamefont{Brida}},
  \bibinfo{author}{\bibfnamefont{M.}~\bibnamefont{Maiuri}},
  \bibinfo{author}{\bibfnamefont{G.}~\bibnamefont{Cerullo}},
  \bibinfo{author}{\bibfnamefont{H.}~\bibnamefont{Schramm}},
  \bibinfo{author}{\bibfnamefont{J.}~\bibnamefont{Christoffers}},
  \bibnamefont{et~al.}, \bibinfo{journal}{Nat Commun}
  \textbf{\bibinfo{volume}{4}}, \bibinfo{pages}{1602} (\bibinfo{year}{2013}),
  \urlprefix\url{http://dx.doi.org/10.1038/ncomms2603}.

\bibitem[{\citenamefont{Karabunarliev and
  Bittner}(2003{\natexlab{a}})}]{karabunarliev:3988}
\bibinfo{author}{\bibfnamefont{S.}~\bibnamefont{Karabunarliev}}
  \bibnamefont{and} \bibinfo{author}{\bibfnamefont{E.~R.}
  \bibnamefont{Bittner}}, \bibinfo{journal}{The Journal of Chemical Physics}
  \textbf{\bibinfo{volume}{119}}, \bibinfo{pages}{3988}
  (\bibinfo{year}{2003}{\natexlab{a}}),
  \urlprefix\url{http://link.aip.org/link/?JCP/119/3988/1}.

\bibitem[{\citenamefont{Karabunarliev and
  Bittner}(2003{\natexlab{b}})}]{karabunarliev:4291}
\bibinfo{author}{\bibfnamefont{S.}~\bibnamefont{Karabunarliev}}
  \bibnamefont{and} \bibinfo{author}{\bibfnamefont{E.~R.}
  \bibnamefont{Bittner}}, \bibinfo{journal}{The Journal of Chemical Physics}
  \textbf{\bibinfo{volume}{118}}, \bibinfo{pages}{4291}
  (\bibinfo{year}{2003}{\natexlab{b}}),
  \urlprefix\url{http://link.aip.org/link/?JCP/118/4291/1}.

\bibitem[{\citenamefont{Karabunarliev and Bittner}(2004)}]{karabunarliev:10219}
\bibinfo{author}{\bibfnamefont{S.}~\bibnamefont{Karabunarliev}}
  \bibnamefont{and} \bibinfo{author}{\bibfnamefont{E.~R.}
  \bibnamefont{Bittner}}, \bibinfo{journal}{Journal of Physical Chemistry B}
  \textbf{\bibinfo{volume}{108}}, \bibinfo{pages}{10219}
  (\bibinfo{year}{2004}), \urlprefix\url{http://dx.doi.org/10.1021/jp036587w}.

\bibitem[{\citenamefont{Karabunarliev and
  Bittner}(2003{\natexlab{c}})}]{karabunarliev:057402}
\bibinfo{author}{\bibfnamefont{S.}~\bibnamefont{Karabunarliev}}
  \bibnamefont{and} \bibinfo{author}{\bibfnamefont{E.~R.}
  \bibnamefont{Bittner}}, \bibinfo{journal}{Physical Review Letters}
  \textbf{\bibinfo{volume}{90}}, \bibinfo{eid}{057402}
  (pages~\bibinfo{numpages}{4}) (\bibinfo{year}{2003}{\natexlab{c}}),
  \urlprefix\url{http://link.aps.org/abstract/PRL/v90/e057402}.

\bibitem[{\citenamefont{Noriega et~al.}(2013)\citenamefont{Noriega, Rivnay,
  Vandewal, Koch, Stingelin, Smith, Toney, and Salleo}}]{Noriega:2013uq}
\bibinfo{author}{\bibfnamefont{R.}~\bibnamefont{Noriega}},
  \bibinfo{author}{\bibfnamefont{J.}~\bibnamefont{Rivnay}},
  \bibinfo{author}{\bibfnamefont{K.}~\bibnamefont{Vandewal}},
  \bibinfo{author}{\bibfnamefont{F.~P.~V.} \bibnamefont{Koch}},
  \bibinfo{author}{\bibfnamefont{N.}~\bibnamefont{Stingelin}},
  \bibinfo{author}{\bibfnamefont{P.}~\bibnamefont{Smith}},
  \bibinfo{author}{\bibfnamefont{M.~F.} \bibnamefont{Toney}}, \bibnamefont{and}
  \bibinfo{author}{\bibfnamefont{A.}~\bibnamefont{Salleo}},
  \bibinfo{journal}{Nat Mater} \textbf{\bibinfo{volume}{advance online
  publication}},  (\bibinfo{year}{2013}),
  \urlprefix\url{http://dx.doi.org/10.1038/nmat3722}.

\bibitem[{\citenamefont{Gao et~al.}(2010)\citenamefont{Gao, Martin, Niles,
  Wise, Thomas, and Grey}}]{doi:10.1021/jp104111h}
\bibinfo{author}{\bibfnamefont{Y.}~\bibnamefont{Gao}},
  \bibinfo{author}{\bibfnamefont{T.~P.} \bibnamefont{Martin}},
  \bibinfo{author}{\bibfnamefont{E.~T.} \bibnamefont{Niles}},
  \bibinfo{author}{\bibfnamefont{A.~J.} \bibnamefont{Wise}},
  \bibinfo{author}{\bibfnamefont{A.~K.} \bibnamefont{Thomas}},
  \bibnamefont{and} \bibinfo{author}{\bibfnamefont{J.~K.} \bibnamefont{Grey}},
  \bibinfo{journal}{The Journal of Physical Chemistry C}
  \textbf{\bibinfo{volume}{114}}, \bibinfo{pages}{15121}
  (\bibinfo{year}{2010}),
  \eprint{http://pubs.acs.org/doi/pdf/10.1021/jp104111h},
  \urlprefix\url{http://pubs.acs.org/doi/abs/10.1021/jp104111h}.

\bibitem[{\citenamefont{Zimmerman et~al.}(2012)\citenamefont{Zimmerman, Xiao,
  Renshaw, Wang, Diev, Thompson, and Forrest}}]{doi:10.1021/nl302172w}
\bibinfo{author}{\bibfnamefont{J.~D.} \bibnamefont{Zimmerman}},
  \bibinfo{author}{\bibfnamefont{X.}~\bibnamefont{Xiao}},
  \bibinfo{author}{\bibfnamefont{C.~K.} \bibnamefont{Renshaw}},
  \bibinfo{author}{\bibfnamefont{S.}~\bibnamefont{Wang}},
  \bibinfo{author}{\bibfnamefont{V.~V.} \bibnamefont{Diev}},
  \bibinfo{author}{\bibfnamefont{M.~E.} \bibnamefont{Thompson}},
  \bibnamefont{and} \bibinfo{author}{\bibfnamefont{S.~R.}
  \bibnamefont{Forrest}}, \bibinfo{journal}{Nano Letters}
  \textbf{\bibinfo{volume}{12}}, \bibinfo{pages}{4366} (\bibinfo{year}{2012}),
  \eprint{http://pubs.acs.org/doi/pdf/10.1021/nl302172w},
  \urlprefix\url{http://pubs.acs.org/doi/abs/10.1021/nl302172w}.

\bibitem[{\citenamefont{Pensack et~al.}(2012)\citenamefont{Pensack, Guo,
  Vakhshouri, Gomez, and Asbury}}]{doi:10.1021/jp2083133}
\bibinfo{author}{\bibfnamefont{R.~D.} \bibnamefont{Pensack}},
  \bibinfo{author}{\bibfnamefont{C.}~\bibnamefont{Guo}},
  \bibinfo{author}{\bibfnamefont{K.}~\bibnamefont{Vakhshouri}},
  \bibinfo{author}{\bibfnamefont{E.~D.} \bibnamefont{Gomez}}, \bibnamefont{and}
  \bibinfo{author}{\bibfnamefont{J.~B.} \bibnamefont{Asbury}},
  \bibinfo{journal}{The Journal of Physical Chemistry C}
  \textbf{\bibinfo{volume}{116}}, \bibinfo{pages}{4824} (\bibinfo{year}{2012}),
  \eprint{http://pubs.acs.org/doi/pdf/10.1021/jp2083133},
  \urlprefix\url{http://pubs.acs.org/doi/abs/10.1021/jp2083133}.

\bibitem[{\citenamefont{Bakulin et~al.}(2012)\citenamefont{Bakulin, Rao,
  Pavelyev, van Loosdrecht, Pshenichnikov, Niedzialek, Cornil, Beljonne, and
  Friend}}]{Bakulin16032012}
\bibinfo{author}{\bibfnamefont{A.~A.} \bibnamefont{Bakulin}},
  \bibinfo{author}{\bibfnamefont{A.}~\bibnamefont{Rao}},
  \bibinfo{author}{\bibfnamefont{V.~G.} \bibnamefont{Pavelyev}},
  \bibinfo{author}{\bibfnamefont{P.~H.~M.} \bibnamefont{van Loosdrecht}},
  \bibinfo{author}{\bibfnamefont{M.~S.} \bibnamefont{Pshenichnikov}},
  \bibinfo{author}{\bibfnamefont{D.}~\bibnamefont{Niedzialek}},
  \bibinfo{author}{\bibfnamefont{J.}~\bibnamefont{Cornil}},
  \bibinfo{author}{\bibfnamefont{D.}~\bibnamefont{Beljonne}}, \bibnamefont{and}
  \bibinfo{author}{\bibfnamefont{R.~H.} \bibnamefont{Friend}},
  \bibinfo{journal}{Science} \textbf{\bibinfo{volume}{335}},
  \bibinfo{pages}{1340} (\bibinfo{year}{2012}),
  \eprint{http://www.sciencemag.org/content/335/6074/1340.full.pdf},
  \urlprefix\url{http://www.sciencemag.org/content/335/6074/1340.abstract}.

\bibitem[{\citenamefont{Tamura et~al.}(2006)\citenamefont{Tamura, Bittner, and
  Burghardt}}]{Tamura:2006}
\bibinfo{author}{\bibfnamefont{H.}~\bibnamefont{Tamura}},
  \bibinfo{author}{\bibfnamefont{E.~R.} \bibnamefont{Bittner}},
  \bibnamefont{and}
  \bibinfo{author}{\bibfnamefont{I.}~\bibnamefont{Burghardt}},
  \bibinfo{journal}{J. Chem. Phys.}  (\bibinfo{year}{2006}),
  \urlprefix\url{http://arxiv.org/cond-mat/abs/0610790}.

\bibitem[{\citenamefont{Tamura et~al.}(2007)\citenamefont{Tamura, Bittner, and
  Burghardt}}]{tamura:021103}
\bibinfo{author}{\bibfnamefont{H.}~\bibnamefont{Tamura}},
  \bibinfo{author}{\bibfnamefont{E.~R.} \bibnamefont{Bittner}},
  \bibnamefont{and}
  \bibinfo{author}{\bibfnamefont{I.}~\bibnamefont{Burghardt}},
  \bibinfo{journal}{The Journal of Chemical Physics}
  \textbf{\bibinfo{volume}{126}}, \bibinfo{eid}{021103}
  (pages~\bibinfo{numpages}{5}) (\bibinfo{year}{2007}),
  \urlprefix\url{http://link.aip.org/link/?JCP/126/021103/1}.

\bibitem[{\citenamefont{Tamura et~al.}(2008)\citenamefont{Tamura, Ramon,
  Bittner, and Burghardt}}]{tamura:107402}
\bibinfo{author}{\bibfnamefont{H.}~\bibnamefont{Tamura}},
  \bibinfo{author}{\bibfnamefont{J.~G.~S.} \bibnamefont{Ramon}},
  \bibinfo{author}{\bibfnamefont{E.~R.} \bibnamefont{Bittner}},
  \bibnamefont{and}
  \bibinfo{author}{\bibfnamefont{I.}~\bibnamefont{Burghardt}},
  \bibinfo{journal}{Physical Review Letters} \textbf{\bibinfo{volume}{100}},
  \bibinfo{eid}{107402} (pages~\bibinfo{numpages}{4}) (\bibinfo{year}{2008}),
  \urlprefix\url{http://link.aps.org/abstract/PRL/v100/e107402}.

\bibitem[{\citenamefont{Bittner et~al.}(2005)\citenamefont{Bittner,
  Karabunarliev, and Ye}}]{bittner:034707}
\bibinfo{author}{\bibfnamefont{E.~R.} \bibnamefont{Bittner}},
  \bibinfo{author}{\bibfnamefont{S.}~\bibnamefont{Karabunarliev}},
  \bibnamefont{and} \bibinfo{author}{\bibfnamefont{A.}~\bibnamefont{Ye}},
  \bibinfo{journal}{The Journal of Chemical Physics}
  \textbf{\bibinfo{volume}{122}}, \bibinfo{eid}{034707}
  (pages~\bibinfo{numpages}{13}) (\bibinfo{year}{2005}),
  \urlprefix\url{http://link.aip.org/link/?JCP/122/034707/1}.

\bibitem[{\citenamefont{Peumans et~al.}(2003)\citenamefont{Peumans, Uchida, and
  Forrest}}]{peumans:158}
\bibinfo{author}{\bibfnamefont{P.}~\bibnamefont{Peumans}},
  \bibinfo{author}{\bibfnamefont{S.}~\bibnamefont{Uchida}}, \bibnamefont{and}
  \bibinfo{author}{\bibfnamefont{S.~R.} \bibnamefont{Forrest}},
  \bibinfo{journal}{Nature} \textbf{\bibinfo{volume}{425}},
  \bibinfo{pages}{158} (\bibinfo{year}{2003}),
  \urlprefix\url{http://dx.doi.org/10.1038/nature01949}.

\bibitem[{\citenamefont{Pereverzev and Bittner}(2006)}]{pereverzev:104906}
\bibinfo{author}{\bibfnamefont{A.}~\bibnamefont{Pereverzev}} \bibnamefont{and}
  \bibinfo{author}{\bibfnamefont{E.~R.} \bibnamefont{Bittner}},
  \bibinfo{journal}{The Journal of Chemical Physics}
  \textbf{\bibinfo{volume}{125}}, \bibinfo{eid}{104906}
  (pages~\bibinfo{numpages}{7}) (\bibinfo{year}{2006}),
  \urlprefix\url{http://link.aip.org/link/?JCP/125/104906/1}.

\bibitem[{\citenamefont{Bittner and Silva}(2013)}]{Bittner:2013aa}
\bibinfo{author}{\bibfnamefont{E.~R.} \bibnamefont{Bittner}} \bibnamefont{and}
  \bibinfo{author}{\bibfnamefont{C.}~\bibnamefont{Silva}},
  \bibinfo{journal}{Nature Communications} \textbf{\bibinfo{volume}{submitted}}
  (\bibinfo{year}{2013}).

\end{thebibliography}

\end{document}